# Strain Controlling Catalytic Efficiency of Water Oxidation for $Ni_{1-x}Fe_xOOH$ alloy

*Ester Korkus Hamal and Maytal Caspary Toroker[*]*

Department of Materials Science and Engineering, Technion - Israel Institute of Technology, Haifa 3200003, Israel

**Abstract**

A catalyst surface may be exposed to strain due to application of load or interfacing with a substrate with large lattice mismatch. In order to test the effect of strain on catalytic efficiency, we use Density Functional Theory +U (DFT+U) to model water oxidation on expanded and contracted surfaces of the $Ni_{1-x}Fe_xOOH$ alloy, one of the best known catalysts for water splitting. We find that a low amount of iron content has a similar effect as of applying compressive strain. Due to high oxidation state of $Fe^{4+}$ at the active site, the Fe-O bond is shorter than in pure FeOOH, which is beneficial for extracting electrons from states delocalized on Fe and Ni atoms. At 33% of Fe content the efficiency is even better since $Fe^{3+}$ is at the active site and can easily change oxidation state during the reaction. However, the efficiency drops at higher iron percentages since the surface is unstable and may form FeOOH aggregates. We find that the best performance is obtained at 33% iron content and 5% expansion. Therefore, in addition to changing iron content, strain can also be used as a control handle for improving water splitting catalysis.

[*]Corresponding author: E-mail: maytalc@technion.ac.il , Tel.: +972 4 8294298.

# 1. Introduction

Water splitting has attracted great interest in recent years due to its potential of generating energy without causing pollution.[1] Further advancement in this technology is an important challenge due to the low performance or stability of available catalysts.[2] Hence, a top priority is to design better catalysts for water splitting.[3-4]

One of the best candidates is nickel oxyhydroxide with iron content ($Ni_{1-x}Fe_xOOH$) that has excellent efficiency at alkaline conditions and is now studied widely.[5-9] [10]An outstanding example is utilizing the $Ni_{1-x}Fe_xOOH$ alloy as a catalyst atop of hematite photoanode.[11-13] Another example is a layered $BiVO_4/FeOOH/NiOOH$ photoanode.[14]

Iron content is essential since without this component the material's catalytic activity is very poor.[15-16] Many experimental and theoretical studies have been devoted to understand why iron improves performance.[17-19] Several insights have been gained. The main observation was that iron is the active site. Furthermore, iron can have several oxidation states which facilitates chemical activity.

Hence, altering material composition through doping or alloying is a natural route to engineer better catalysts. External conditions such as the application of load or interfacing with a substrate can also have an effect on catalytic performance. However, using strain as a design strategy for improving catalysts has been less explored.[20-21]

In this work we consider the effect of strain on water oxidation with $Ni_{1-x}Fe_xOOH$. Our approach is to use Density Functional Theory +U (DFT+U) in order to model water oxidation while the surface is contracted/expanded. This study extends beyond previous studies by adding a detailed analysis of iron composition and location dependence under strain. We find that both iron concentration and strain affect the local bond distances near the active site and have a direct consequence on efficiency.

# 2. Methods and Calculation Details

The VASP program was used for spin-polarized Density Functional Theory (DFT) calculations.[22-23] The chosen functional is the Perdew-Burke-Ernzerhof (PBE)[24] functional with the DFT+U formalism of Duradev et al.[25] at an effective U-J term of 5.5 eV and 3.3 eV for Ni and Fe as previously done for doped NiOOH.[17, 26-29] We found in our previous work,[30] that these values for U-J are essential for capturing the qualitative experimental result that Fe doping reduces the overpotential required for water oxidation. In this study, we also provide results showing that using a different functional such as Heyd-Scuseria-Ernzerhof (HSE06)[31] fails to agree with experiment on the favorable effect of Fe doping on catalysis. We also show that van der Waals D2 corrections of Grimme[32] have negligible contribution (see Tables S1-S4). Projected-

augmented wave (PAW) potentials replaced the electrons of Ni 1s2s2p3s3p, Fe 1s2s2p3s3p, and O 1s.[33-34]

The unit cell of β-NiOOH[17, 35-38] was cleaved at the (0$\bar{1}$5) plane since this surface was studied in previous literature and we wanted to simplify the analysis and compare to previous results.[17] The facet (001) has been recently suggested[16, 39] due to its surface stability, but according to our band edge calculations then this facet should be less active.[28] Hence, the same slab sizes were built as in refs. [[17, 28, 40]]. The iron content was considered to be with at% of: 8%, 25%, 33%, and 42% at various locations of nickel atomic substitution. We considered 8% as in previous studies,[18] and higher concentrations below 50% where the material is known to be catalytically more efficient. The Fe content in $Ni_{1-x}Fe_xOOH$ alloys was considered at several representative locations in the surface unit cell, while the selection of the models was based on varying number of iron atoms close and away from the active site (see Figure 1). At 8% both iron and nickel were considered at the active site for comparison, but at alloying concentration of 25% and above we considered iron as the active site since this site was found chemically active in previous studies.[10, 17] The final configuration that was selected for applying strain was the one with the lowest total energy and overpotential at preliminary calculation tests (Tables 5S-8S). Hence, for 8%, 25%, 33%, and 42% Fe contents, "location 2", "location 2", "location 3", and "location 2" were selected, respectively. The energy cutoff of 600 eV and k-point Gamma-centered grid of 2x2x1 were converged to within <1 meV. The ion positions were converged until the force components on all ions were less than 0.03 eV/Å. Additional details are given in the supporting information and a data set collection of computational results is available in the ioChem-BD repository.[41]

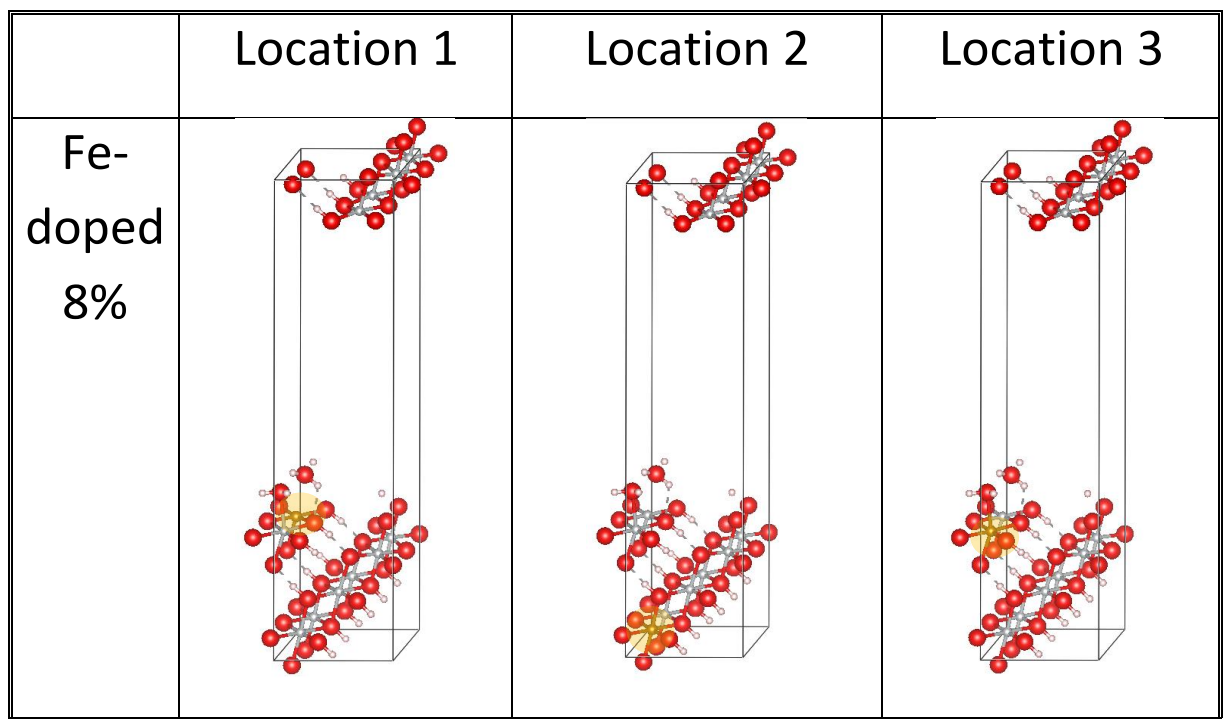

| | | | |
|---|---|---|---|
| Fe-doped 25% | 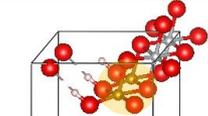 | 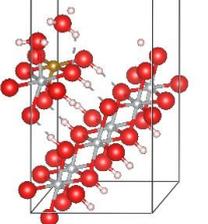 | 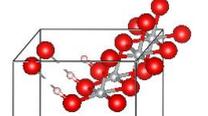 |
| Fe-doped 33% | 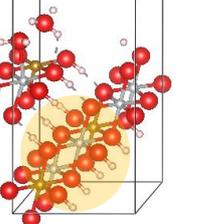 | 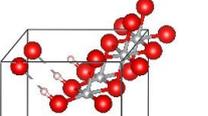 | 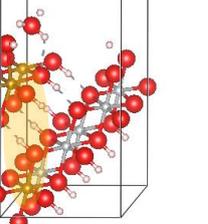 |
| Fe-doped 42% | 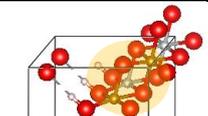 | 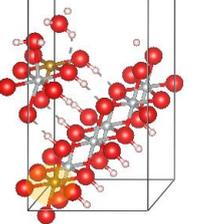 | 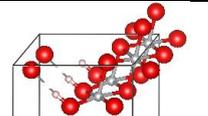 |

**Figure 1.** Surface unit cells of $Ni_{1-x}Fe_xOOH$ alloys for different Fe percentages and locations. The cells are shown at the first reaction intermediate, where water adsorbes on the surface prior to deprotonation. Red, gray, and gold spheres represent O, Ni, and Fe atoms, respectively. The positions of the varying Fe iron atoms are highlighted by the orange shaded area.

We considered the following four-step mechanism that has been studied in ref. [17], for comparison, at alkaline conditions (see Figure 2):[18]

$$A + h^+ + OH^- \rightarrow B + H_2O \quad (1)$$

$$B + h^+ + OH^- \rightarrow C \quad (2)$$

$$C + h^+ + OH^- \rightarrow D + H_2O \quad (3)$$

$$D + h^+ + OH^- \rightarrow A + O_2 \quad (4)$$

where intermediate "A" is a slab with adsorbed water molecules, intermediate "B" has one hydroxyl group termination, intermediate "C" is the same as intermediate "A" but has an additional oxygen atom penetrating the surface, and intermediate "D" is similar to intermediate "B" but the additional oxygen atom penetrating the surface. The free energy for the reactions was calculated by adding the previously reported Zero Point Energy (ZPE) corrections and entropic contributions of pure β-NiOOH since these additions were shown stable with variations in composition.[17, 42] The free energies were calculated at the operating conditions for the OER: 1 Volts and pH=14, by adding a constant energy -eU for the applied voltage and a term of $-k_BT \cdot ln10 \cdot pH$ for the *pH* where $k_B$ is Boltzmann constant and $T$ is the temperature of 298.15 K. The free energies without *pH* and *T* corrections are provided in the supporting information (Tables 9S-13S). The overpotential is defined as voltage needed to add to the calculated electrochemical potential so that all reaction free energies are negative. The intermediate reaction cells are expanded or contracted significantly (up to 10%) in order to be able to see the effect of strain. The expansion or contraction was performed by elongating or contracting the surface lattice vectors and allowing only the ions to optimize in their positions. The lattice vectors were changed only in the directions of the $\vec{a}$ and $\vec{b}$ vectors. For example, the expansion is reflected by the distance between the sheets elongated from 2.624 Å to 2.722 Å (see Figure 2).

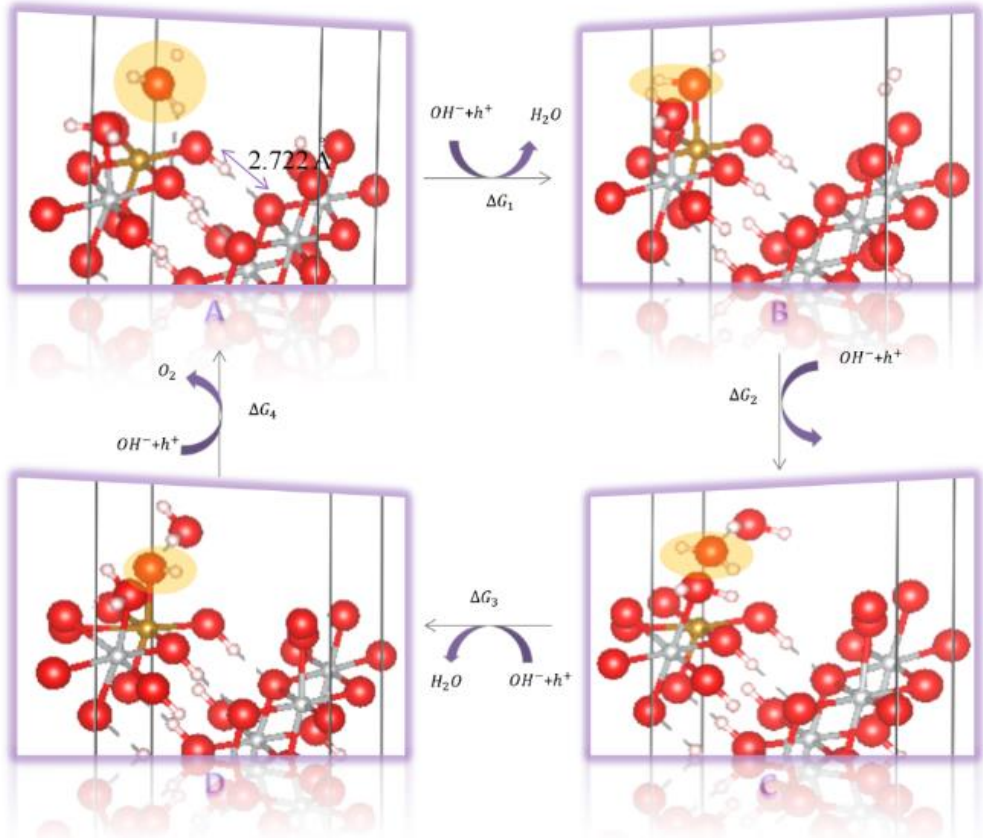

**Figure 2.** Catalytic cycle of water oxidation for the Ni$_{1-x}$Fe$_x$OOH alloy at x=8% and 5% expansion. Red, gray, and gold spheres represent O, Ni, and Fe atoms, respectively. Created with VESTA visualization package.[43]

## 3. Results

In this section, we present free energy calculations of the water oxidation reaction for the Ni$_{1-x}$Fe$_x$OOH alloy under applied strain. First, we focus on the pure NiOOH case and analyze the effect of strain on each intermediate reaction of water oxidation. Next, we show for 8 at% Fe-doping the effect of both doping and strain. We then extend the analysis to higher Fe contents and discuss how both alloying percentage and strain affect the overpotential required for water oxidation.

The overpotential required for water oxidation on pure NiOOH is 0.61 eV (see Table 1). The process determining the overpotential is the chemical reaction with the highest free energy, which is the first deprotonation step for pure NiOOH (-0.11 eV in Table 1). In contrast, with the application of 10% contraction, the highest free energy belongs to the second reaction step. As seen in Figure 3, the most dominant change is for reactions 2 and 4, where oxygen penetrates and leaves the surface, respectively. The changes for reaction 2 and 4 are dramatic and can span over 2.5 eV difference when comparing 10% contraction vs. 10% expansion (see Table 1). Expansion makes more space for an oxygen atom to penetrate, but disfavors oxygen release. Therefore, the reaction 2 free energy reduces with expansion and the reaction 4 free energy increases with expansion.

The rest of the reactions 1 and 3 are less sensitive to strain for pure NiOOH, although the effect is significant enough to reduce the overpotential at 5% contraction or expansion (as seen in Table 1, ~0.1 eV decrease in the overpotential and free energy of the first reaction upon expansion). A correlation between the free energies of reactions 1 and 3 and the application of strain can be identified while fixing the fractional coordinates of the atoms. In this case, there is a general linear relation between the free energies of reaction 1 and 3 and strain, that is there is a general increase in the overpotential when expansion is applied. This expansion increases the interatomic distance, which localizes the atomic orbitals, and inhibits the release of localized electrons during deprotonation reactions.

**Table 1.** Free energies and overpotentials for pure NiOOH with or without strain at pH=14 and V=1 Volts. Units are eV.

| Reaction | -10% | -5% | 0 | 5% | 10% |
|---|---|---|---|---|---|
| A to B | -0.18 | -0.23 | -0.11 | -0.19 | 0.04 |
| B to C | 0.36 | -0.24 | -0.71 | -1.12 | -1.70 |
| C to D | -0.41 | -0.36 | -0.40 | -0.51 | -0.44 |
| D to A | -2.57 | -1.97 | -1.58 | -0.98 | -0.71 |
| Overpotential | 1.08 | 0.49 | 0.61 | 0.52 | 0.76 |

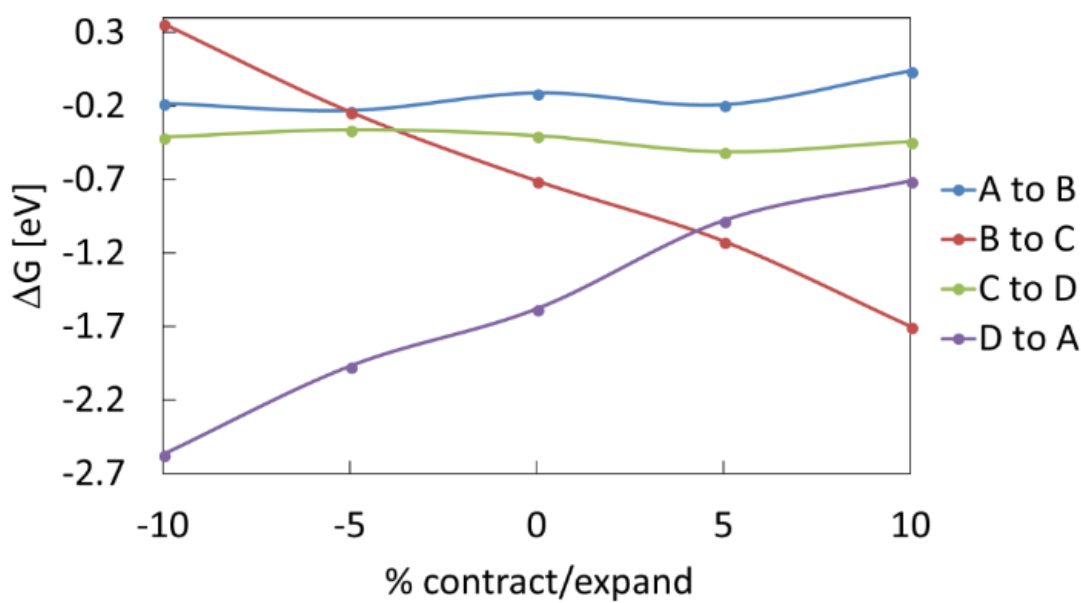

**Figure 3.** Free energies for water oxidation reaction for pure NiOOH as a function of % contraction/expansion.

Similarly to pure NiOOH, the 8% Fe-doped case also has an overpotential that is determined by the first two reactions (see Table 2 and Figure 4). However, as seen in Table 2, the intersection of the reaction 1 and 2 free energies occurs at zero compression (the free energies for the first two reactions are ~-0.4 eV), since the bond distances around the active site are shorter for the doped case. The average Fe-O bond distance at the active site is 1.912 Å, which is shorter than the 1.952 Å average Ni-O distance at the active site of pure NiOOH. The Fe-O bonds are even shorter than typical Fe-O bonds of pure FeOOH where it is expected that the distances would be larger than pure NiOOH according to the order of Ni and Fe atoms in the periodic table of elements (Ni has a larger atomic number and therefore smaller ionic radius). The short Fe-O bonds at the active site result from a high +4 oxidation state of iron at the active site of 8% Fe doped NiOOH (calculated magnetization for intermediate A is 3.6 Bohr magneton).[18] The calculated shorter bond distances are in good agreement with previous experiment.[10] And evidence for the presence of $Fe^{4+}$ and $Fe^{3+}$ has been reported in various studies.[44] The deprotonation reaction is affected by the bond distances and determines the overpotential. Hence, *the key to the overpotential decrease upon Fe doping is the similarity of the free energies of reactions 1 and 2. Our testbed of applying strain helps to explain that the overpotential lowers upon doping since Fe decreases the bond distances near the active site and contributes to less compression needed to equalize the overpotentials of reactions 1 and 2*. Moreover, the positive effect of reducing the overpotentil of pure NiOOH upon Fe-doping is generally maintained also under strain (compare overpotentials in Tables 1 and 2). Furthermore, the overpotential decreases by an additional 0.12 eV upon the application of 5% strain (Table 2).

**Table 2.** Free energies and overpotentials for Fe-doped NiOOH (8%) with or without strain at pH=14 and V=1 Volts. Units are eV.

| Reaction | -10% | -5% | 0 | 5% | 10% |
|---|---|---|---|---|---|
| A to B | -0.79 | -0.81 | -0.41 | -0.47 | -0.57 |
| B to C | 0.76 | 0.25 | -0.35 | -0.73 | -1.23 |
| C to D | -0.82 | -0.92 | -0.69 | -0.71 | -0.59 |
| D to A | -1.96 | -1.33 | -1.36 | -0.90 | -0.42 |
| Overpotential | 1.48 | 0.97 | 0.37 | 0.25 | 0.30 |

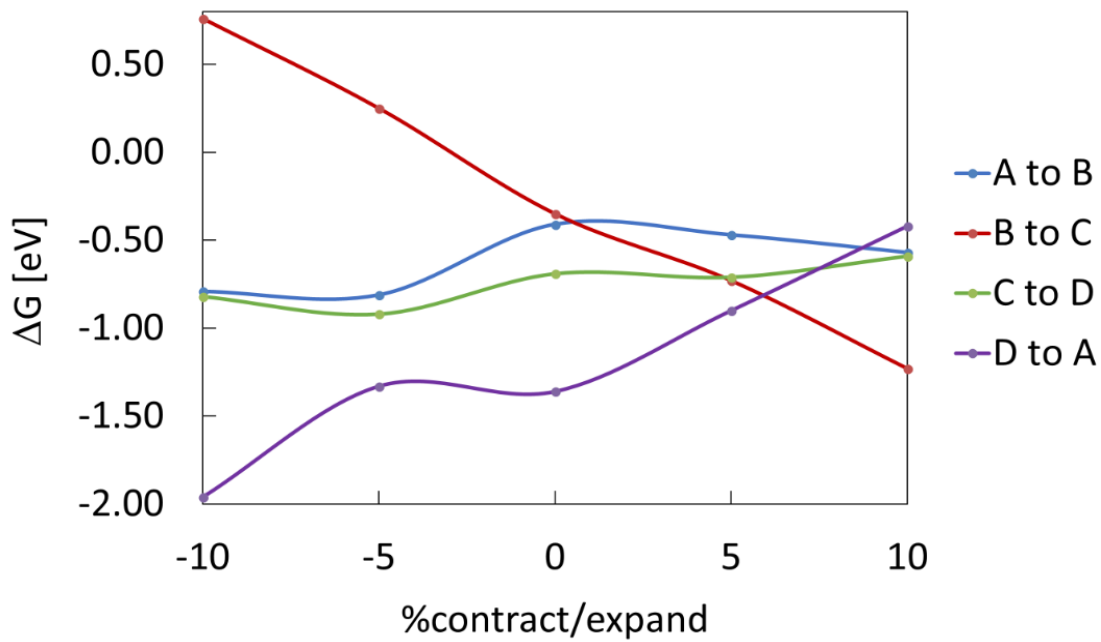

**Figure 4.** Free energies for water oxidation reaction for Fe-doped NiOOH (8%) as a function of % contraction/expansion.

The overpotential further decreases with the increase of Fe content till it reaches an optimal efficiency at 33% Fe concentration (see Table 3 and Figure 5). The overpotential for Fe content in the range of 25% to 42% is similar (~0.3 eV) with a slight preference for 33%. The free energy of reactions 1 and 2 are very similar at all Fe percentages (see Table 3). At 33% Fe content, the preferred (lowest energy) configuration includes Fe at the active site as well as neighboring to the active site. This preference of Fe clustering at the active site results in a +3 oxidation state for Fe at the active site (corresponds to a calculated atomic magnetization of 4.1 Bohr magneton), which is similar to the oxidation state of Fe in pure FeOOH. The small oxidation state of Fe at 33% causes a reduction in overpotential since the ionization of $Fe^{3+}$ is easier than that of $Fe^{4+}$. This electronic contribution is important and persists even when the slab model is partially optimized in geometry (see Figure 4). The surrounding Ni atoms are also critical for the activity since the Fe states are hybridized (delocalized) with Ni states at all Fe concentrations (see Figure 6) and this enables deprotonation. As seen in Figure 6, the hybridization is dominant in the valence band where there are chemically active states that loose an electron for deprotonation. The small oxidation state of Fe corresponds to longer chemical bonds (see Table 4 and Figure 7. As seen in Table 4, all of the bonds around the active site are the longest at 33% with an average length of 1.97 Å, which is longer than in the pure NiOOH case and corresponds to the expected ordering of Fe and Ni elements in the periodic table of elements.

As a side note, we remark that not all bonds react to the contraction or expansion in a trivial way, that is, for example, bonds to not necessarily elongate as a result of 5% expansion (as seen in Table 4). Another important location where bond distances are central to the catalytic efficiency is at the location where oxygen penetrates (see Table 14S and Figure 5S). There, expect for some exceptions, the bond distances vary with expansion or contraction but do not change significantly as a result of Fe content increase, probably due to similar chemical environment around this site.

At high iron contents of 42%, the atomic location of Fe atoms close to the active site results in a low overpotential and is less stable (see higher energy of intermediate A in Table 8S). Indeed, the creation of inactive FeOOH aggregates have been observed experimentally.[10] Hence, the preferred (low energy) configuration has less Fe atoms near the surface and $Fe^{4+}$ at the active site (as seen in Table 15S, the atomic magnetization of Fe is 3.6 Bohr magneton at intermediate A for 8%, 25%, and 42% of Fe content) and a higher overpotential compared to 33% iron content (see Figure 5).

**Table 3.** Free energies and overpotentials for $Ni_{1-x}Fe_xOOH$ alloy at different iron concentrations with or without strain at pH=14 and V=1 Volts. Units are eV.

| Reaction | Pure NiOOH | Fe-doped 8% | Fe-doped 25% | Fe-doped 33% | Fe-doped 42% |
|---|---|---|---|---|---|
| A to B | -0.11 | -0.41 | -0.45 | -0.51 | -0.57 |
| B to C | -0.72 | -0.35 | -0.43 | -0.46 | -0.42 |
| C to D | -0.40 | -0.69 | -0.56 | -0.63 | -0.66 |
| D to A | -1.59 | -1.36 | -1.38 | -1.21 | -1.16 |
| **Overpotential[eV]** | 0.61 | 0.37 | 0.29 | 0.26 | 0.30 |

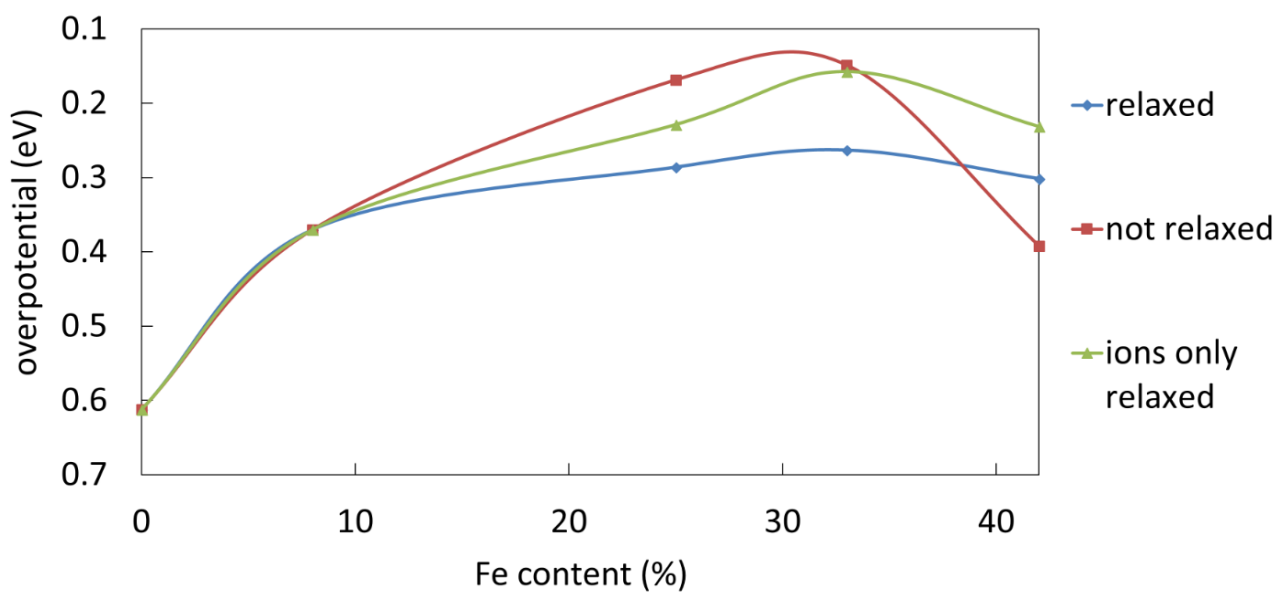

**Figure 5:** Overpotential for water oxidation with mixed Ni-Fe oxyhydroxide catalyst as a function of Fe content. For Fe content above 8%, the initial geometry was taken as

8% Fe content. Red = Fixed geometry, blue = fully relaxed geometry (program keyword ISIF=3), and green = lattice constants fixed and ions are relaxed in their positions (program keyword ISIF=2).

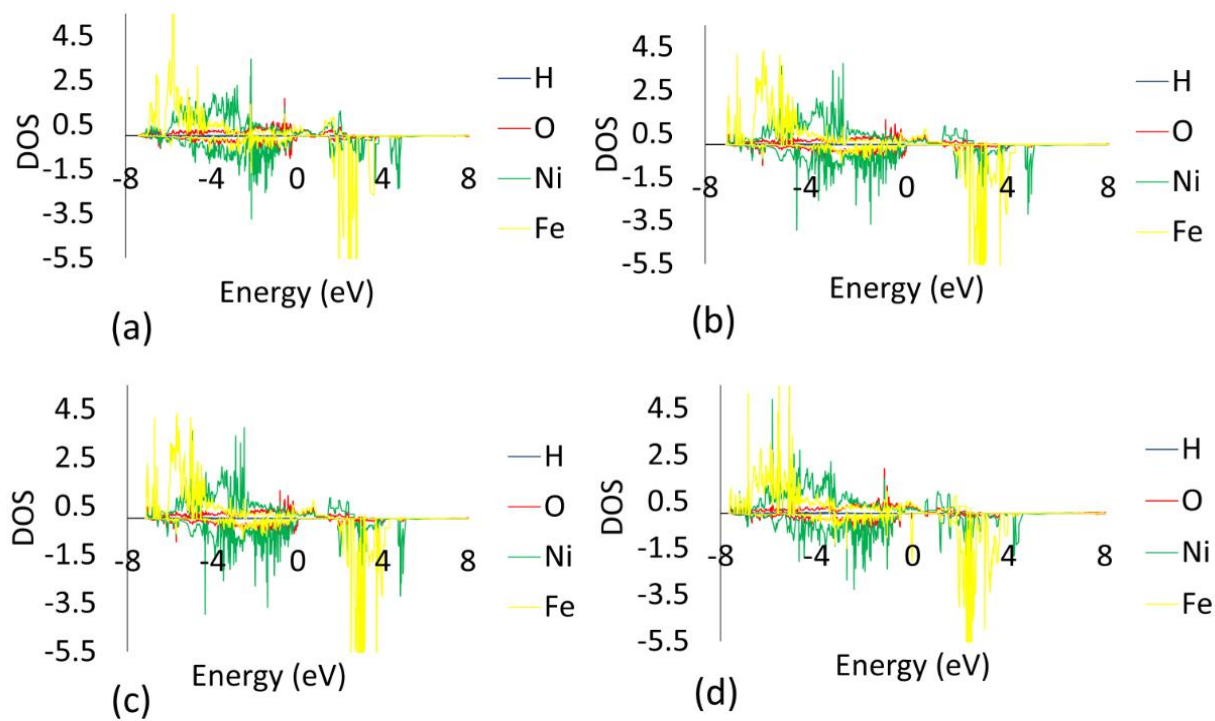

**Figure 6.** Density of states for the $Ni_{1-x}Fe_xOOH$ alloy at a) x=8%, b) x=25%, c) 33%, and d) 42%.

|  | no strain | | | | |
| --- | --- | --- | --- | --- | --- |
| % | 8 | 25 | 33 | 42 | PURE |
| Fe-O (1) | 1.873 | 1.868 | 1.964 | 1.874 | 1.937 |
| Fe-O (2) | 1.921 | 1.914 | 2.004 | 1.921 | 1.941 |
| Fe-O (3) | 1.891 | 1.884 | 1.967 | 1.877 | 1.925 |
| Fe-O (4) | 2.001 | 1.990 | 2.009 | 2.038 | 2.012 |
| Fe-O (5) | 1.875 | 1.867 | 1.893 | 1.843 | 1.947 |
| average | 1.912 | 1.905 | 1.967 | 1.911 | 1.952 |
|  | 5% strain | | | | |
| % | 8 | 25 | 33 | 42 | PURE |
| Fe-O (1) | 1.886 | 1.875 | 1.832 | 1.984 | 1.950 |
| Fe-O (2) | 1.958 | 1.938 | 1.959 | 1.942 | 1.970 |
| Fe-O (3) | 1.920 | 1.908 | 1.961 | 1.914 | 1.946 |
| Fe-O (4) | 1.981 | 1.982 | 1.940 | 2.003 | 2.013 |
| Fe-O (5) | 1.885 | 1.879 | 1.902 | 1.775 | 1.965 |
| average | 1.926 | 1.916 | 1.919 | 1.924 | 1.969 |
|  | -5% strain | | | | |
| % | 8 | 25 | 33 | 42 | PURE |
| Fe-O (1) | 1.935 | 1.822 | 1.830 | 1.879 | 1.893 |
| Fe-O (2) | 1.937 | 1.901 | 1.924 | 1.824 | 1.895 |
| Fe-O (3) | 1.919 | 1.909 | 1.822 | 1.923 | 1.884 |
| Fe-O (4) | 2.076 | 2.038 | 2.028 | 2.032 | 2.002 |
| Fe-O (5) | 1.911 | 1.830 | 1.827 | 1.846 | 1.900 |
| average | 1.956 | 1.900 | 1.886 | 1.901 | 1.915 |

**Table 4.** Bond lengths around the active site for different Fe percentages at no applied strain. The Fe site is octahedrally coordinated, but one bond is absent since the site is located at the surface. The average of the five remaining bonds is indicated. Chemical bond numbers are shown in Figure 7. Units are eV.

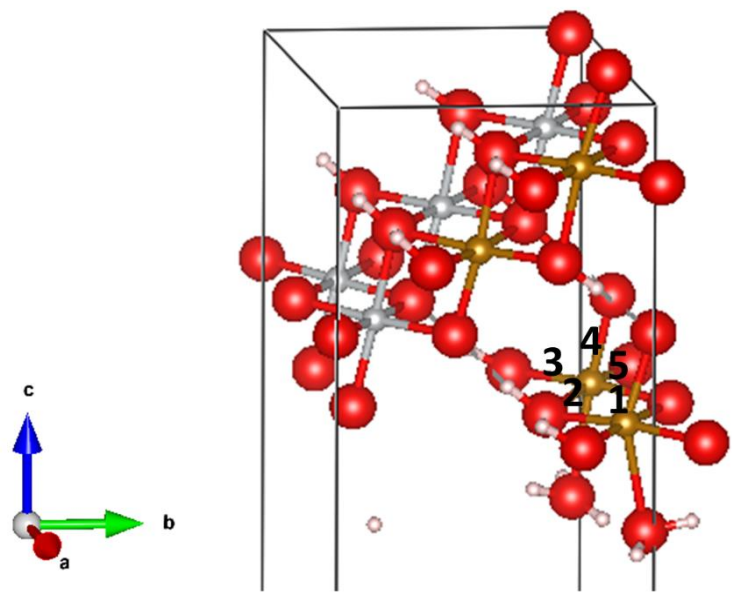

**Figure 7:** Atomic arrangement at the active site. The structure is relaxed for 8% iron doping. The bonds surrounding the Fe active site are numbered. Red, gray, and gold spheres represent O, Ni, and Fe atoms, respectively. Created with VESTA visualization package.[43]

The best performance is achieved when applying 5% expansion strain to NiOOH with 33% Fe metallic content. As seen in Figure 8, strain has a significant effect on the overpotential for all percentages of Fe in NiOOH. In all cases, expansion is desired in order to reach optimal hybridization of all atomic orbitals, enabling further delocalization and easier extraction of charge during deprotonation.

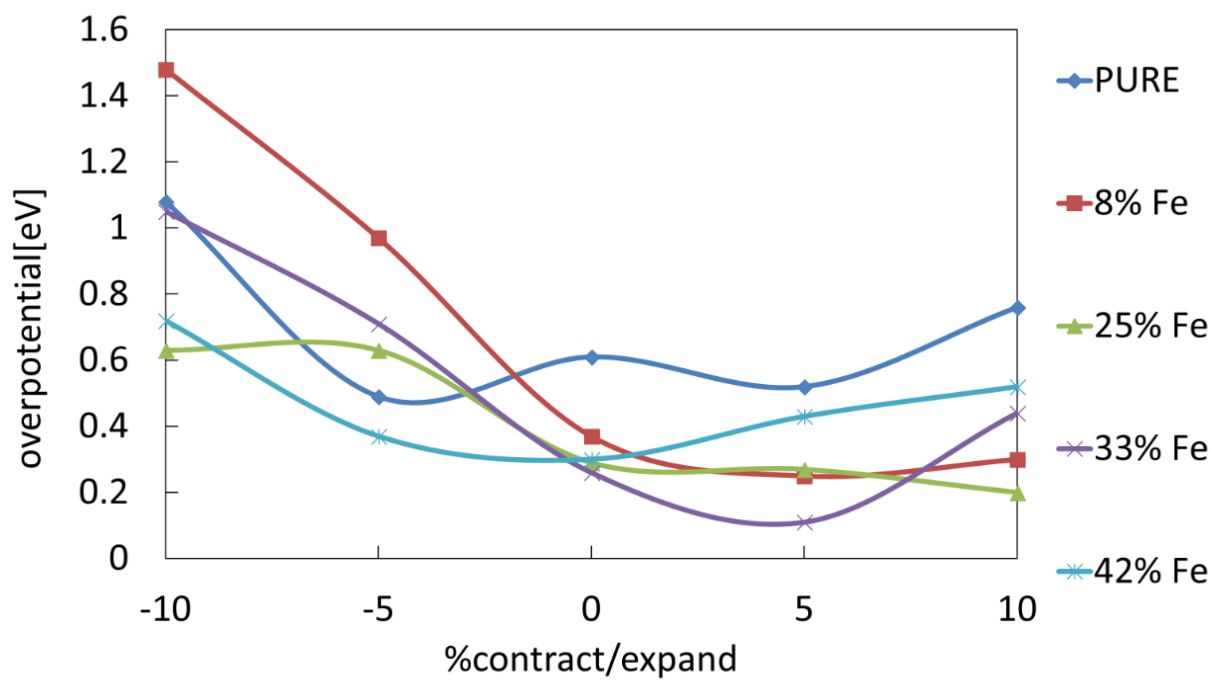

**Figure 8:** Overpotential vs. strain for varying Fe alloying percentages.

## 4. Conclusions

This research contributes to understanding the outstanding catalytic activity of $Ni_{1-x}Fe_xOOH$, one of the best heterogeneous catalysts for water oxidation under alkaline conditions. We perform an analysis based on DFT+U calculations of the free energies required for intermediate reactions of water oxidation while applying compression or expansion strain. Our analysis of applying strain at various Fe contents helps explain the role of both Fe and Ni atoms in the strong catalytic ability of $Ni_{1-x}Fe_xOOH$.

The catalytic efficiency of $Ni_{1-x}Fe_xOOH$ has four important contributions:

1. Fe can acquire several oxidation states which is beneficial for the oxidation process. We find that in most Fe contents then the active site's oxidation state is +4, but at 33% of Fe then some Fe atoms cluster close to the active site and the preferred oxidation state is +3 as in pure FeOOH. In the +3 oxidation state then Fe needs less energy to oxidize and the overpotential is lowest.
2. The Fe and Ni states occupied states are hybridized and allow better charge extraction during deprotonation. We find this is present at all Fe contents and at all strains applied according to the projected density of states. Hence, the presence of Ni atoms surrounding the active site is vital for good performance.
3. The bond distances at the active site need to be optimal for the first deprotonation reaction to be as low as the second intermediate reaction step. For pure NiOOH, the first two reactions have equal free energies (but high overpotential) at -5% compressive strain. For $Ni_{1-x}Fe_xOOH$ with $Fe^{4+}$ at the active site, the high oxidation state of Fe reduces bond distances at the active site and compressive strain is not needed to reduce the overpotential.

4. High Fe contents located on the surface at 42% alloying are less stable (higher energy). According to previous experiments[10] this results in the formation of FeOOH aggregates that reduce performance.

Moreover, at all Fe content percentages, application of strain affected the overpotential. Strain has a direct effect on the distances and hybridization of the bonds at the active site. Hence, applying strain is another useful control handle that can be used to understand and optimize catalytic efficiency.

## Acknowledgements

This research was supported by the Nancy and Stephen Grand Technion Energy Program (GTEP), the I-CORE Program of the Planning and Budgeting Committee, The Israel Science Foundation (Grant No. 152/11), the SPIRA Fund for Applied Research in the Field of Energy, and a grant from the Ministry of Science and Technology (MOST), Israel. This work was supported by the post LinkSCEEM-2 project, funded by the European Commission under the 7th Framework Programme through Capacities Research Infrastructure, INFRA-2010-1.2.3 Virtual Research Communities, Combination of Collaborative Project and Coordination and Support Actions (CP-CSA) under grant agreement no RI-261600.

## Supporting information available

Further details on the calculated free energies and on the unit cells used for the surfaces are provided in the supporting information and in the ioChem-BD repository.

## Table of Contents Graphic

The influence of mechanical strain on water oxidation catalysis for nickel oxyhydroxide at different levels of iron content.